# Computational Social Science and Critical Studies of Education and Technology: An Improbable Combination?

Rebecca Eynon[a]* and Nabeel Gillani[b]

[a]*Oxford Internet Institute and Department of Education, University of Oxford, Oxford, UK;* [b]*College of Arts, Media and Design and D'Amore-McKim School of Business, Northeastern University, Boston, MA, USA*

*Rebecca Eynon, Oxford Internet Institute, University of Oxford, 1 St Giles, Oxford, OX1 3JS. Email: rebecca.eynon@oii.ox.ac.uk

# Computational Social Science and Critical Studies of Education and Technology: An Improbable Combination?

As belief around the potential of computational social science grows, fuelled by recent advances in machine learning, data scientists are ostensibly becoming the new experts in education. Scholars engaged in critical studies of education and technology have sought to interrogate the growing datafication of education yet tend not to use computational methods as part of this response. In this paper, we discuss the feasibility and desirability of the use of computational approaches as part of a critical research agenda. Presenting and reflecting upon two examples of projects that use computational methods in education to explore questions of equity and justice, we suggest that such approaches might help expand the capacity of critical researchers to highlight existing inequalities, make visible possible approaches for beginning to address such inequalities, and engage marginalised communities in designing and ultimately deploying these possibilities. Drawing upon work within the fields of Critical Data Studies and Science and Technology Studies, we further reflect on the two cases to discuss the possibilities and challenges of reimagining computational methods for critical research in education and technology, focusing on six areas of consideration: criticality, philosophy, inclusivity, context, classification, and responsibility.

Keywords: computational social science; digital education; justice; AI; research methods; equity; big data

**Introduction**

The proliferation of 'big data'—the very fine-grained, real-time data generated when people interact with digital technologies—together with advances in algorithms and the hardware that supports computation, has enabled significant developments in machine learning and 'AI'. Digital trace data is now routinely created, collected and analysed across all levels and settings of education, through the proliferation of digital technologies that are used for all aspects of learning, teaching and administration; and this adds to the ever-growing collection of administrative data in education by governments and intergovernmental organisations. Proponents argue that the use of such data and associated machine learning techniques is the best way of understanding the world, enabling "a data revolution and a radical reimagining of teaching and learning" (OECD, 2021:n.p.). AI, as with big data before it, attracts a celebratory and magical status (Crawford, et al., 2014, Suchman, 2023) that is "symbolic of a wider rhetoric and imaginary that is used to garner support and spread their role and adoption" (Kitchen, 2014: xvi).

This excitement and belief around the powers of computational approaches for education has not occurred by accident, but has rather come about largely through the discursive and economic activities of the commercial sector who benefit directly from these activities (Ball, 2021; Grimaldi and Ball, 2021; Miglani and Burch, 2021; Williamson, 2017; Komljenovic et al., 2023). It is further supported by several related but distinct academic communities, with Educational Data Mining (EDM), Learning Analytics (LA) and Artificial Intelligence in Education (AIED) being the most prominent.

These groups, both within and outside of academia, are united by two broad characteristics. First, most of their work aligns with a knowledge tradition that can be described as the "New Science of Education" (Furlong and Whitty, 2017; Eynon, 2023). This is a view heavily supported within policy circles internationally, and typically located within a positivist ontology, where a focus on learning outcomes and measuring the efficacy of learning interventions using methodologies from the natural sciences is paramount (Furlong and Whitty, 2017). Second, these communities tend to primarily focus on learning, typically from a psychological perspective that seeks to "optimiz[e] cognitive learning processes" (Mayer, 2012), instead of focusing on sociological questions that draw attention to issues of education, and questions of power, justice and social change (Gillani and Eynon, 2023; Knox et al., 2020).

Significantly, researchers within data science communities are now viewed as the central experts in education by policy makers, even in cases where data scientists have a very limited background in this field (Kitchen, 2014; Williamson, 2017, 2021). This is a crucial problem. As Boellstroff notes, "knowledge production is never separate from the knowledge producer" (Boellstorff, 2013: n.p.). This means that the framing of the problem to be solved, the selection of data, the creation of models from that data, and the uses to which these models are put, are all very much shaped by the expertise and perspectives of those who build them (Crawford, 2021). Data scientists are likely to be expert in data collection, management and analysis, but they often come from other fields such as engineering or physics with limited (if any) training or academic experience of education. Yet, they are increasingly the ones who get to set the agenda. Those with experience and expertise in education, particularly those who take a critical perspective, have been effectively side-lined in current thinking and policy.

The principal response to this ever-growing "datafication" of education has been the development of a nuanced and growing area of study that directly critiques this direction of travel. Such work, which often draws on fields such as Critical Data Studies (CDS) and Science and Technology Studies (STS), includes a focus on: the problems of the data and the logic of the approaches themselves (Castillo and Gillborn, 2023; Kizilcec and Lee, 2022); the wider implications of the use of such data and associated models in various contexts of education and governance (Eynon, 2022; Grimaldi and Ball, 2021; Nemorin, 2024; Selwyn, 2019; Williamson, et al., 2022); the promotion of the use of data by local communities and grassroots organisations to make visible the injustice they are experiencing and to support their call for change (Berland and Garcia, 2024; Sriprakash et al., 2024); and a focus on regulation and governance (Lindh and Nolin, 2016; Livingstone et al., 2024).

However, despite this rich research, there has been very limited debate about how computational approaches might be used as one part of the multifaceted critical response to the datafication of education—and what the risks or benefits of their use might be. There has also been limited discussion about the value such computational methods might have in the broader field of critical studies of education and technology. There are exceptions. For example, responsible learning analytics (Cerratto-Pargman et al., 2023) has emerged as one response—although as the name suggests, this work tends to focus on learning and not education; the growing "QuantCrit" community, which draws on Critical Race Theory (CRT) to interrogate and improve quantitative approaches in social science that originated in the field of education (Castillo and Gillborn, 2023); and the occasional use of large scale social media data to explore critical educational questions (Rosenberg et al., 2024). However, it is fair to say that the debate on the value of computational methods in field of critical studies of education and technology has been relatively limited. This can be contrasted, for example, with debates in sociology, where almost two decades ago, academics were asked to reflect on the very future of the discipline (Burrows and Savage, 2007; 2014), with evolutions of this debate continuing today (Diaz-Bone et al., 2020).

**Aims and objectives**

In this paper, we wish to discuss the feasibility and desirability of using computational approaches for critical studies of education and technology. Over the past few years there has been a growing call for work that could be described as computational social science for education (Gillani and Eynon, 2023) or educational data science (McFarland et al., 2021). Computational social science is typically defined as "the development and application of computational methods to complex, typically large-scale, human (sometimes simulated) behavioral data" (Lazer et al., 2020: 1060), where computational methods are typically drawn from computer science, engineering or physics. Similarly, Educational Data Science can be thought of as "an umbrella [term] for a range of new and often nontraditional quantitative methods (such as machine learning, network analysis, and natural language processing) applied to educational problems often using novel data" (McFarland et al., 2021:2). Here, we use the term "computational social science for education" to better connect with the intention to use these computational methods as part of social science research, and with the wider discussions on computational social science that is occurring across disciplines. Within this landscape, we are interested primarily in the use of computational methods which utilise recent advances in machine learning and other data science techniques, typically using digital trace data and large-scale data sets; yet acknowledge that such work may also at times encompass more "traditional" quantitative approaches (e.g. regression analysis).

In what follows, we wish to explore not only whether such computational methods can be used to explore questions of interest to scholars focused on critical studies of education and technology, but also whether such data and approaches should be used to ask such questions at all.

The reasons for considering this are threefold. First, a focus on computational approaches may offer alternative data sources and methods that can potentially lead to substantive insights that may further a critical agenda in education. Second, these approaches may enable critical researchers to more directly enact positive social change via engagement with policy makers and data scientists, using methodological approaches that both groups value. In other words, "talking the same language" as the current "experts" in the datafication of education—or at least one similar enough to potentially shift the current debates and promote more positive social change. Third, more in-depth engagement with computational approaches may offer new forms of critique and ways to challenge the use and application of such approaches in learning and education.

Alongside these potential possibilities, are of course significant challenges. It is, for example, difficult if not impossible to remove the objectivist and behaviourist assumptions that are implicitly built into many machine learning models in use today (Knox et al., 2020), particularly as foundation models (i.e. "models that are trained on broad data at scale and are adaptable to a wide range of downstream tasks" (Bommasani et al., 2021)) become ever more prevalent. Likewise, the extractive, biased and oppressive forms of data collection that form the basis of many of these models (Birhane et al., 2021; Crawford, 2021)—and also the small number of commercial actors in this domain who seek to further profit and intensify their power in education (Komljenovic et al., 2023)—are far removed from the values and beliefs of the majority of researchers who are committed to a critical agenda in education.

While the current context is sobering, nothing is inevitable; it is possible to develop viable alternatives. The aim of this article is to contribute to a debate in education of how to shift the balance of power in who shapes the methodological and research agenda back from the private to the public sphere, and help reimagine how machine learning and other associated techniques might potentially be designed and used in ways that advance democratic aims and social justice. In the discussion below, we offer a reflexive review of two case studies to explore some of the complexities and possibilities involved in reimagining computational techniques in this way.

**The two cases: school review sites and school boundaries**

There are many questions that critical scholars engaged in researching digital technologies and education can ask (Macgilchrist, 2021). Here, we focus on questions of equity and justice, and suggest that computational methods might contribute to this agenda in three ways: 1) making visible existing inequalities ("highlighting inequalities"); 2) highlighting possible approaches to addressing such inequalities ("identifying possibilities"); and 3) engaging communities, particularly those at the margins, in developing and ultimately deploying these possibilities ("participatory design").

In this paper, we present and reflect on two cases of projects that use computational approaches in education that speak to these three objectives. The first case, based on research on school ratings/reviews websites, offers an example of using computational approaches to highlight and map inequalities in education systems ("highlighting inequalities"). The second case, based on research on school boundaries, offers an example of using computational approaches as part of participatory, justice-orientated research ("identifying possibilities" and "participatory design"). We argue that both examples represent two broad types of computational social science for education, and that each might offer insights into questions of importance to critical scholars of education and technology. Additionally, these examples might also spark reflection upon the complexities of using computational methods, and the feasibility of reimagining such methods for critical research in education and technology.

*Case one: school review sites*

This first case demonstrates how computational approaches might be used to make inequalities visible in education systems. To do this, we use an example of a paper (authored by us and a wider research team) that contributed to the debates on the use of school review websites to support school choice by parents and guardians. The aim of the study was to explore the subjective assessments of quality made by parents / guardians as evidenced by their written reviews on school ratings websites, and the relationship between these assessments and measures of school quality (Gillani et al.,

2021).

To do this, the research team combined three different data sets. First, a web scraper was used to collect over 800,000 reviews posted by parents for more than 110,000 schools on a U.S. K–12 school ratings site, with permission from the platform. These 110,000 schools were then linked to the Stanford Educational Data Archive (SEDA) (Reardon et al., 2019) on two performance metrics: (1) average test scores, which provide a snapshot-in-time measure of student performance, and (2) student learning rates, which indicate how much student cohorts improve on standardized tests each year, and which was used in the study as the measure of school quality. Finally, the school addresses available on the ratings site were geocoded and linked to tract-level estimates of race, socioeconomic status, and other demographics provided by the 2010 Census and 2015 American Community Survey (Gillani et al., 2021).

Combining these data sets, and using advances in natural language processing (NLP) to analyse the text of these reviews, the study found that: (1) schools in urban areas and those serving more affluent families were more likely to receive reviews, (2) review language correlated with standardized test scores but not with student learning rates, and (3) the language of the reviews indicated several racial and income-based disparities in K–12 education. For example, phrases about "the pta" (Parent Teacher Association) and "emails" were negatively correlated with the percentage of students receiving free or reduced lunch; and parents whose children were at schools with a higher proportion of white students were distinguished by reviews containing the phrase "small school" (Gillani et al., 2021).

Overall, the findings suggested that parents who read school reviews may be accessing, and making decisions based on, biased perspectives that reinforce achievement gaps; which undermines the policy intention of using school ratings websites to support school choice in a way that proponents suggest will reduce existing inequalities in education (Gillani et al., 2021). For example, the fact that parents' reviews related to standardized test scores (which track race and income) but did not significantly relate to student progress scores (which may better reflect how well teachers and staff at the school may be helping students learn over time) threatens to perpetuate a tendency for parents to choose schools based on test scores and demographics instead of factors that correlate with student learning, that risks performative responses from schools. In addition, parents from certain racial or income groups may read disproportionately more about certain topics, like the PTA or regular email communication with teachers, which in turn could shape what they expect, or even demand, from schools (Gillani et al., 2021).

*Mapping inequity within the status quo*

The above paper, we suggest, is an example of the responsible use of large-scale computational analysis to highlight inequalities in educational systems.

It is responsible in that it embodies several principles that would be largely supported by proponents of algorithmic fairness (Kizilcec and Lee, 2022). A central issue for such computational approaches are well established concerns around data privacy and informed consent, and risks of bias. In this case although data was scraped as part of the analysis, this was public data that was designed to be used to obtain and communicate parents' views. Second, to analyse the words and phrases from parents' reviews the team chose to fine-tune the pre-trained, deep neural network language model BERT (Devlin et al., 2019), which was pre-trained on a large corpus of English books and English Wikipedia (Wolf et al., 2020). While this corpus, like any, will be subject to biases, which has implications for BERT, it is reasonable to argue that the data on which BERT is based is significantly less extractive than other models commonly seen in this space (Srnicek, 2017; Crawford, 2021).

The paper further documents the process of the collection, selection and curation of data used in the study as well as justifies the fine-tuning of BERT. It is also cautious about delineating what can and cannot be claimed from the results (Suresh and Guttag, 2020; Baker and Hawn, 2022; Kizilcec and Lee, 2022). For example, although BERT is a black box model, the study team used a variety of techniques to try to make the research as transparent as possible, incorporating both computational approaches to model transparency along with human sense checking throughout, as documented in the paper (Gillani et al., 2021).

Still, the paper also illustrates the limits of algorithmic fairness as a framework for critically assessing the use of computational methods to understand social systems. The paper tends not to disrupt the status quo in the way the research is framed, nor does it fully attend to the structural constraints and power relations that underpin the findings (Eynon, 2024). For example, school choice, a highly contested area (e.g. Ball, 1993) and one reflecting the continued marketisation of education, is taken as a given in the study. The paper presents both sides of the argument for school choice, however, the conclusion reached is that, "whether choice-based systems hold promise for bridging inequalities in education or not, they are increasingly being deployed, and their effectiveness is predicated on families being well-informed and well-supported to decide which schools are best for their children" (Gillani et al., 2021:1). Given this, and the rise of popularity in school ratings sites as resources to help parents research and identify schools (Lovenheim and Walsh, 2017), the study team argues that they are an important focus of research. This stance is reasonable but lacks the critical stance familiar to readers of *Learning, Media and Technology*.

These challenges continue in relation to questions of categorisation and measurement. The study takes a measure of school "quality" that focuses on how well a school helps students make academic progress (as measured by test scores) over time. This risks narrowing the questions asked about a complex social and educational issue to only what can be measured (Kizilcec and Lee, 2022; Eynon, 2022). The narrowness and problems of this measure are acknowledged in the paper, though ultimately is was used in part because some members of the team viewed it as a valid measure; in part because it is used by policy makers, and also because a sensible measure needed to be found. Nevertheless, this choice of measure does have broader implications for the kinds of knowledge produced about education (Hakimi et al., 2021). For computational social science to be of value to critical scholars, there is a need to question the logics of the measures used, and to find or campaign for alternatives (Berland and Garcia, 2024).

These issues are handled reflexively in the paper, and the findings are valuable as they can be used to counter assumptions about the value of digital platforms in supporting school choice policy. Given the nature and scale of the school reviews data, it would be challenging to gain a comprehensive view of what was happening on the review site without using computational methods, and also to highlight the relationships between these activities and the nature of the schools being reviewed. As such, we suggest there is an important place for this kind of work, within a broader set of methodological approaches used in critical studies of education and technology. Yet, the study also demonstrates the limits of this kind of research, and the paper is somewhat unconvincing in terms of its potential to reclaim computational methods for critical education research.

*Case two: Community-driven, machine-informed rezoning to foster more integrated schools*

This second case is an example of studies that use machine learning to support participatory research. Action based, participatory work with marginalised communities is considered to be fundamental to a meaningful response to the datafication of everyday life; and the use of computational methods can potentially extend the existing response in education as in other areas of social life (Costanza-Chock, 2020; D'Ignazio and Klein, 2019).

The research in this case study is being carried out by a team from two universities in the US, and is currently in progress. The case focuses on the issue of school segregation. In the US, student access to quality teachers, academic resources, and networks that might help students access equitable opportunities to improve their quality of life tracks race and income (Reardon, et al., 2018). More diverse and integrated schools may help reduce such inequalities (Chetty et al. 2022), but integration remains elusive across many districts. One reason is because of how districts draw "school attendance boundaries", or catchment areas that determine which students attend which schools. School districts in the US can draw attendance boundaries however they like—including in ways that would advance an agenda of desegregation. However, changing attendance boundaries can trigger significant community backlash, and such concerns often dominate and impede changes that would enable more equitable, less-segregated boundaries (Billingham and Hunt, 2016; Gillani et al., 2023a).

This case study aims to advance equity-promoting boundaries in school districts. At present, many districts use database systems such as ArcGIS, often informed by contractors and consultants, to explore and eventually present potential redistricting plans for community input. There are three main challenges with the current approach: 1) a vast number of plans could be generated and examined in theory, yet in practice, creating plans using existing database systems is a time-consuming and cumbersome process (Becker and Solomon, 2020) and often prone to arbitrary decision-making; 2) such plans do not always reveal the space of "what's possible": i.e., if there are boundary configurations that might advance integration aims while perhaps also satisfying other concerns that parents might have (e.g. travel time to school), thereby challenging the "zero-sum" assumptions many hold about access to quality educational environments; and 3) the views of more privileged parents tend to be over represented during the community engagement phases of these consequential policy change processes (Siegel-Hawley et al., 2018; Gillani et al., 2023b).

The researchers in this study have aimed to address these challenges by forming a researcher–practitioner partnership (RPP) with a large public US school district to develop and evaluate redistricting algorithms that draw on computational methods in ways that also factor in community input. They are adopting a value-sensitive design (VSD) approach to this work (Koster et al, 2022), that seeks to identify and factor in stakeholders' values when designing technology. The study also seeks to surface how technology might affect human values, and how those values, in turn, can shape the development of technologies.

In this study, there is an intention to take a three-phase approach. First identifying possibilities by using computational methods to determine which school boundaries have the most potential for change, and therefore, which communities to work with to help support such change ("identifying possibilities"). Second, using the RPP to engage with community organizations and both school and district leadership, to co-design the methods and the associated platform that can be utilised to capture what communities value and prioritize when it comes to deciding which schools their children should attend, which can then be utilised in later modelling of new boundaries through a process of participatory design ("participatory design"). Such methods might include surveys, voice notes, map annotations etc. Third, to use the platform with the community, and use this feedback to inform the algorithm that creates different possible boundary redrawings, transparently showing how the values, priorities, and other community feedback captured in the prior phase influenced their development. The RPP can then analyse this feedback using both qualitative and computational methods and translate the priorities and preferences into factors that are re-inputted into the redistricting algorithm. Using an iterative approach to identifying possibilities with the RPP will ultimately lead to a co-defined proposal for attendance boundary changes. Throughout, the study is actively seeking to privilege and engage with community members from historically marginalized or under-heard groups, for example, by weighting the algorithms and door-to-door canvassing to support engagement (Gillani et al., 2023a; Gillani et al., 2023b).

*Towards a justice-orientated approach*

This second case reflects a commitment to using data sets and models with care and transparency, again employing a number of principles that would be largely supported by proponents of algorithmic fairness (Kizilcec and Lee, 2022), but also trying to bring in stronger commitments to data justice (Costanza-Chock, 2020, Dencik, 2022) in three ways.

First, by making redistricting algorithms interpretable. Families and other stakeholders can see which factors the models weigh to produce alternative boundaries designed to achieve a particular goal (which may help address challenges to transparency and legitimacy), and underlying biases in the data or models can be interrogated by the community, research team and beyond. Second, by viewing these algorithms as a socio-technical system that supports "human+AI" collaborations by factoring in community preferences (Gillani et al., 2023a) to automatically create different boundary scenarios. This may not just be more efficient than the low-tech efforts in the past, but may also encourage more explicit and productive discussions of human biases, prejudice and values. Third, by creating new civic technologies that surface AI-generated policy proposals to help elevate traditionally underrepresented voices by creating new channels for public participation (Mehrabi et al., 2022). These technologies might also help connect families to the experiences of other families, potentially helping to mitigate individualistic thinking that harms the collective.

There are of course significant challenges. Participatory approaches are not a panacea, and the realities of this kind of research are typically highly ethically challenging, not least due to the burden placed on marginalised groups (Pierre et al., 2021). Nevertheless, it creates an important space for debates and research on school rezoning. As with the first case, the team has reflected on the sources of data used. The data used in this system (i.e. geocoded, anonymized counts of the number of students living in each census block that attend each school; alongside the participative qualitative data) is not particularly extractive: it is already collected by school districts as a part of their registration and enrolment processes. There is additional work that could be done to interrogate the categorisation of such data, and questions of identity and how intersectional approaches could be productively used in such processes (Costanza-Chock, 2020; D'Ignazio and Klein, 2019; Berland and Garcia, 2024). Nevertheless, this work illustrates how computational approaches could be used to challenge existing social structures, make visible debates about school zones, and encourage a plurality of voices from marginalised groups (Costanza-Chock, 2020; D'Ignazio and Klein, 2019). This second case then perhaps offers some hope of the possibility of reimagining computational methods for critical studies of education and technology.

**Reimagining computational methods**

The two cases demonstrate in different ways how computational approaches could be potentially valuable in critical studies of education and technology that are focused on questions of equity and justice. In the first case, while not directly changing the status quo (in terms of school choice and the wider social structure) it provides evidence in terms policy makers now value that highlights the problems of current educational systems ("highlighting inequalities"), and offers insights that are difficult to achieve with other methods. In the second case, there is a stronger commitment to changing the status quo through surfacing possible alternatives ("identifying possibilities") using approaches from computational social science, and engaging with communities to promote positive social change ("participatory design"). Like the first case, it is likely to appeal to policy makers through its focus on large scale data, and leads to findings that are not always possible with other methods. But it is also more likely than the first case to facilitate reflexive critique of existing inequities in education, and via the participatory engagement raise questions about the realities and appropriateness of using computational approaches for such an agenda.

In different ways, then, these two cases offer insights into the potential reclaiming of computational methods for critical studies of education and technology. Reflecting on these cases, together with the growing literature in CDS and STS, we identify six areas of focus that may support this reclaiming and reimagining of computational methods as part of a critical agenda in education and technology. These six areas are: criticality, philosophy, inclusivity, context, classification, and responsibility. We consider each in turn.

*Criticality*

We suggest that for computational methods to be of value to the critical education and technology community, researchers need to use theory to frame and shape the research, be reflexive in the questions asked, and contextualize questions for local contexts. These kinds of important questions can be thought of as "critical with a small c" (Selwyn, 2015).

Such studies also need to incorporate a strong commitment to positive social change (Green, 2021), what Selwyn describes as critical with a "capital C", that is, "a desire to foster and support issues of empowerment, equality, social justice and participatory democracy" (Selwyn, 2015: 250). Examples of such big C critical contributions could include a focus on localised change to challenge injustice in education; a call for better data collection and consent practices—allowing people to reclaim their data and regulate how databases are used (e.g. Birhane et al., 2021; Selwyn, 2019); advocate for more ethical and morally accountable use of computational modelling; and contribute, in collaboration with the communities it most impacts, to the surfacing of alternative futures in the use of technology and education (Sriprakash et al., 2024). We suggest that both the cases discussed in this article can be considered as critical with a small c; for example, both work with a specific theory of equity or justice. The second case can additionally be considered as critical with a Big C, given it has a concrete vision of how positive social change might be supported.

*Philosophy*

There is a need to be explicit about the ontological and epistemological commitments of such research, which is not straightforward. As noted above, there are significant philosophical differences between the largely positivist and behaviourist stances utilised in computational approaches, and the typically interpretivist and experiential stances common in critical studies of education and technology. These are not reconcilable, yet making these tensions visible and explicit within the research remains important. Indeed, as in interpretative work, there is a lot of complexity in understanding the varied assumptions and practices that characterise computational research, which potentially lends itself to productive philosophical dialogue between these two different worldviews (Bryant and Raja, 2014; Wagner-Pacifici et al., 2015).

We would suggest a critical realist philosophy of science (e.g. Archer et al., 2016) may be one way forward, as it advocates a realist ontology that many in data science would find appealing, whilst also supporting epistemic relativism and a commitment to social change (Archer, 2020; Clegg, 2016). Others have suggested there could well be promise in drawing on new materialists such as Karen Barad (Giraud et al., 2024; Østerlund et al., 2020). Regardless of the philosophical view taken, what remains important in any work that uses data science as part of a critical agenda is the need to make such tensions and inconsistencies visible.

*Inclusivity*

For any study that aims to speak to questions of equity and justice, there is a need to prioritise marginalised perspectives, promote data practices (both in collection of data and choice of models applied to that data) that reflect on questions of extraction and consent, and consider the power dynamics of the research itself (including the make-up of the research team and who the research is "done to") (Costanza-Chock, 2020; D'Ignazio and Klein, 2019). These issues were important in both the cases discussed above, but perhaps are best exemplified in the second case, given inclusivity was a cornerstone of the entire framing of the project.

The combination of qualitative and computational methods in the second case was also key to supporting inclusion, as the combination of methods helped surface important nuances in parental views on school integration. For example, in the team's work so far with the school district, it has started to identify survey responses and other input from marginalised groups —who prior research suggests may benefit the most from more integrated schools—express concerns about integration. Many of these families are worried about losing access to the specialized support and other programs designed for their children at their current schools. Sometimes, they believe greater resource equity—not necessarily integration—should be the district's aims. This and other findings that buck expected trends have come to light in part because of the research team's "big data" approach to engagement that captured a myriad of voices from all corners of the district. Yet at the same time, using qualitative approaches as well as computational methods was essential for analysing the data, in order to identify the nuances and unique circumstances behind different viewpoints; and to understand them while giving full attention to both individual experience and structural and cultural conditions.

Indeed, those working within the critical sphere typically advocate for the value of using qualitative approaches together with computational methods (e.g. D'Ignazio and Klein, 2020; 2023), despite the challenges of doing so as a result of the philosophical differences described above. Proponents of mixed method approaches highlight how qualitative research is often problematically sidelined or used in the service of large-scale computational approaches. Instead, they suggest, it is important to ask, "how qualitative analysis complicates big data processed through automated means" (Giraud et al., 2024:9); which attends to "the novel dynamics that emerge when knowledges *complicate* rather than *complement* each other" (Giraud et al., 2024:9, drawing on Barad (2007)).

Related to inclusion, a valid critique of the two cases discussed above is that both of these US studies represent a particular worldview which would need to be countered by more work that recognizes colonial power relations (Couldry and Mejias, 2021; Mohamed et al., 2020; Nemorin, 2024). Some may reasonably argue that both cases are still based within a Western, Global North-influenced ontology and epistemology, which are insufficient for justice-orientated research, and other studies of education and technology with a critical orientation (Heath et al., 2023).

*Context*

Context is central, both in the original framing and development of a study (as noted in the discussion on "Criticality" above) and throughout the research process, including in the interpretation of the results. Qualitative approaches are essential to problematise, challenge and interpret themes and findings emerging from computational methods. In the first case discussed above, for example, the research team found that parents of children at more affluent, higher test scoring schools with a higher proportion of white students, were more likely to talk about "special needs" than those posting reviews from schools serving a larger fraction of historically-disadvantaged students. The problem is that keywords such as these are easily stripped of cultural and temporal relations (Giraud et al., 2024) when undergoing analysis with computational methods. Although the research team tried to use approaches that gathered as much context as possible, and the study included multiple instances of "sense checking", there were moments in that case where the research team wished for more qualitative insights. Ultimately, to interpret such data in a meaningful way, requires a qualitative analysis which accounts for theoretical knowledge, and that understands such keywords as just one part of a complex set of socio-technical arrangements (Giraud et al., 2024).

In the case of the keyword "special needs", there is an ongoing debate in education about the extent to which students from historically disadvantaged backgrounds are enrolled in special education programs in US public schools: some argue special education programs serve as another vehicle for segregating these students away from their more affluent counterparts, while others argue that not enough students are receiving the special support they need to succeed (Elder et al., 2021). The empirical finding in the first case—that parents from advantaged schools tend to talk more about "special needs" than others—does not advance this ongoing debate in either direction. However, it does illuminate which types of parents are more or less likely to discuss such a topic in a public online forum, which could form one part of a multifaceted understanding in a future mixed methods study on this topic.

*Classification*

The problems of categorisation and measurement of educational data are well recognised, reflecting philosophical differences in what data "is" (Crawford et al, 2016). This includes debates around how best to classify educational data, the nature of the biases in the data, problematising the neat boundaries that classification encourages, and complicating what the data can be understood to represent (e.g. Berland and Garcia, 2024; Birhane et al., 2021; Diaz-Bone, et al., 2020; Giraud et al., 2024; Madaio et al., 2022).

In both the cases discussed in this article, there was some reflection on the appropriateness of the measures used. However, drawing on more qualitative work that problematises the very notion of categorisation may also have been valuable, for example in considering questions of intersectionality (Hoffmann, 2019), and the underlying challenges of high-stakes assessments used to measure school quality (Madaio et al., 2022).

That said, at times, computational approaches can also contribute to these discussions. For example, using large-scale datasets and computational methods to define, measure, and compare historically difficult-to-define, yet perennially important measures of well-being and opportunity such as "social capital" (Chetty et al., 2022) has been a productive approach.

*Responsibility*

As noted above there is a significant literature on the importance of algorithmic fairness and responsible data science practices, which encompass the need to address issues of fairness, accountability, transparency, and interpretability (Stark et al., 2021). A great deal of this guidance is based on technical approaches to addressing these problems (Stark et al., 2021). These are important for researchers to be aware of and to apply as appropriate, yet ultimately, there is a need to interpret these issues in a broader ethical and moral sense, where researchers also take on a social responsibility for their uses of computational approaches (Nissenbaum, 1996, Cooper et al., 2022; Sahlgren, 2021).

These technical and social responsibilities require a reflexive approach at all stages of the research process, from documenting the process of data collection and analysis, alongside the recognition of problematic data, detailing decisions around the selection and curation of data and the algorithms used, and questioning of the validity of the models and the interpretations of results. This is also needed in the pragmatic choices that contribute in small but important ways to the environmental concerns of such approaches (Selwyn, 2021) and to a "digital degrowth" agenda (Selwyn, 2024). For example, by selecting models that do a sufficient job, as opposed to using the biggest or most "advanced" model, or opting to use local models instead of those deployed in the cloud.

Precisely what such responsibility might look like may varies depending on the study and is very difficult to achieve fully. Yet there is a growing literature on which researchers can draw (e.g. Gebru et al., 2021; Mitchell et al., 2019). In both of the cases we have discussed, responsibility was an important goal. While challenging, researchers, similarly to tech developers, should take responsibility for their practices and the outputs they produce when using computational approaches.

**Conclusion**

In this paper, we have put forward our experiences of using computational approaches to examining questions of equity and justice in education. Although only two cases have been discussed, we hope they contribute to a discussion about if and how, scholars engaged in critical research interrogating the use of digital technologies in education might use computational methods.

On balance, we argue that the use of computational social science is an activity worth pursuing. Echoing other critical work in and beyond education in which scholars envision contexts where computational approaches can be used alongside other methods—finding ways to work with communities to support autonomous data practices that support meaningful local change, and using data and related approaches in a reflexive way, recognising its partial, biased, constructed and often extractive nature (e.g. Dencik, 2022; D'Ignazio and Klein, 2020; 2023; Pangrazio et al., 2024)—we argue that through research that attends to issues of criticality, philosophy, inclusivity, context, classification, and responsibility, computational methods could potentially be valuable in critical studies of education and technology.

For some critical scholars, the arguments we have put forward in this paper may not be sufficiently compelling: the risks of bias, of surveillance, oppression and instrumentalism when using computational methods may be too great. Does the motivation of developing a critical stance and use of such methods just play into data centric, neo-liberal arguments, and further consolidate the power of the commercial sector and particular groups in society? Will, as well recognised by critical scholars working within related domains of data justice and data activism (e.g. Crooks and Currie, 2021; Dencik et al., 2019), any kind of critical agenda be co-opted into a more instrumental one? In our hopes of using such approaches not only to engage on the same playing field as other "policy experts" but also to develop the field methodologically and theoretically, are we being duped by the "magic" of AI? And how can we properly account for the use of such approaches as concerns around the environmental costs of computational methods grow (Selwyn, 2021, 2024)?

While acknowledging these important issues, it is striking how the precise ways that computational approaches can be used as part of a more critical and justice-orientated agenda have rarely been delineated within the field of education. These debates seem ever more pressing as 'AI' becomes written in to research policy at institutional and sector-wide levels in many countries. We hope this article sparks some of these conversations, to encourage critical scholars of education and technology to consider that if these cases do not go far enough, what would a critical computational social science for studies of education and technology look like, and is it something to be desired?